\documentclass[a4paper,11pt]{article}
\usepackage{jinstpub}
\usepackage{lineno}

\usepackage{float}
\usepackage{graphicx}
\usepackage{subcaption}
\usepackage{tikz}
\usetikzlibrary{arrows.meta, positioning}
\usetikzlibrary{calc}
\usepackage{makecell} 
\usepackage{hyperref}

\title{Timing characterization of MALTA and MALTA2 pixel detectors using Micro X-ray source}

\author[1]{G. Dash}
\author[2]{P. Allport}
\author[3]{I. Asensi Tortajada}
\author[1]{P. Behera}
\author[4]{D.V. Berlea}
\author[5]{D. Bortoletto}
\author[6]{C. Buttar}
\author[3]{V. Dao}
\author[4]{L. Fasselt}
\author[3]{L. Flores Sanz de Acedo}
\author[5]{M. Gazi}
\author[7]{L. Gonella}
\author[8]{V. Gonzalez}
\author[3]{G. Gustavino}
\author[3]{S. Haberl}
\author[3]{T. Inada}
\author[1]{P. Jana}
\author[2]{L. Li}
\author[3]{H. Pernegger}
\author[3]{P. Riedler}
\author[3]{W. Snoeys}
\author[3]{C.A Solans Sanchez}
\author[3]{M. van Rijnbach}
\author[3,8]{M. Vazquez Nunez}
\author[1]{A. Vijay}
\author[3]{J. Weick}
\author[4]{S. Worm}

\affiliation[1]{Indian Institute of Technology Madras, Chennai, India}
\affiliation[2]{University of Birmingham, Birmingham, United Kingdom}
\affiliation[3]{CERN, Geneva, Switzerland}
\affiliation[4]{DESY, Zeuthen, Germany}
\affiliation[5]{University of Oxford, Oxford, UK}
\affiliation[6]{University of Glasgow, Glasgow, United Kingdom}
\affiliation[7]{Universit\`{a} degli Studi di Trieste, Trieste, Italy}
\affiliation[8]{Universitat de València, València, Spain}

\emailAdd{ganapati.dash@cern.ch}

\abstract{The MALTA monolithic active pixel detector is developed to address some of the challenges anticipated in future high-energy physics detectors. As part of its characterization, we conducted timing studies necessary to provide a figure of merit for this family of monolithic pixel detectors. MALTA has a metal layer in front-end electronics, and the conventional laser technique is not suitable for timing studies due to the reflection of the laser from the metallic surface. X-rays have been employed as a more effective alternative for penetration through these layers. The triggered X-ray set-up is designed to study timing measurements of monolithic detectors. The timing response  of the X-ray set-up is characterized using an LGAD. The timing response of the MALTA and MALTA2 pixel detectors is studied, and the best response time of MALTA2 pixel detectors is measured at about 2.6 ns.}

\keywords{Particle tracking detectors (Solid-state detectors); Radiation-hard detectors; Solid state detectors}

\begin{document}
\maketitle
\flushbottom

\section{Introduction}

Pixel detectors are crucial in particle physics experiments due to their ability to precisely capture and record the individual locations of particle interactions~\cite{review}. Their high spatial resolution and fine granularity contribute to improved tracking, aid in analyzing complex event topologies, and enhance the overall performance of experiments in unraveling fundamental aspects of particle behavior and interactions. The increasing energy and luminosity of the Large Hadron Collider at CERN~\cite{hllhc,atlaslhc}, along with future experiments, require pixel detectors that can operate reliably in challenging radiation environments while maintaining high detection efficiency, low noise, minimal dead time, and low latency. Therefore, there is a constant effort to design efficient and thinner pixel detectors suitable for high-luminosity collisions. Monolithic silicon pixel detectors~\cite{CMOS,Monopix,darwish,alpide,berda} offer several advantages over traditional bump-based hybrid detectors, including higher precision, lower noise, reduced material usage, and the potential for large-area production at lower cost. Despite their comparatively lower radiation hardness, these benefits have motivated significant R\&D in CMOS-based pixel technologies. Such efforts have led to the development of the MALTA~\cite{penneger2,cardella1} and its second-generation MALTA2~\cite{piro1,milou1} pixel detectors, specifically targeting applications in the HL-LHC upgrades and other future experiments. One important property of pixel detectors is fast timing, defined as the delay or the time taken between the interaction of a particle with the detector and the corresponding detection signal. In this study, a micro X-ray source is designed and employed to investigate the fast-timing performance of the MALTA and MALTA2 detectors.

\section{MALTA and MALTA2 Demonstrator Systems}

MALTA, designed using Tower-Jazz 180~nm technology~\cite{penneger2}, is a robust monolithic active pixel detector resilient to radiation exposure up to a fluence on the order of \(10^{15} \, \text{n/cm}^2\) and a hit rate capability of up to \(100 \, \text{MHz/cm}^2\). Featuring an active area of \(18.3 \, \text{mm}^2\) and a \(512 \times 512\) pixel matrix with \(36.4 \times 36.4 \, \mu\text{m}^2\) pixels, it boasts a quick pixel response time of up to \(20 \, \text{ns}\), meeting the HL-LHC's bunch-crossing identification requirements. The chip's asynchronous read-out eliminates the need for clock distribution, reducing power consumption. MALTA features a double-column read-out architecture and can read out multiple hits per double column. The chip is read out using a Virtex-7 series FPGA. Connection to a VC707 evaluation board is established through an FMC connector as shown in Fig \ref{fig:setup2}, with a specially designed asynchronous oversampling detecting bits from the differential LVDS signals at an effective sampling rate of 1 Gsps, and subsequently read into the control software via IPBUS~\cite{lyre1}.

MALTA2~\cite{vlad,milou2}, a pixel detector demonstrator, is designed using Tower-Jazz 180~nm monolithic CMOS technology, measuring \(20.2 \, \text{mm} \times 10.1 \, \text{mm}\). It is approximately half the size of MALTA, featuring a matrix of \(224 \times 512\) pixels, with each pixel measuring \(36.4 \times 36.4 \, \mu\text{m}^2\). Two process modifications (NGAP and extra deep PWELL/XDPW) have been implemented to increase the charge collection efficiency in the pixel corners~\cite{mod}. It is specifically designed to improve radiation hardness~\cite{MiniMalta} while achieving lower random telegraph signal (RTS) noise in the sensor front-end. Improvements in MALTA2 slow control include using a shift register instead of an Ethernet-like protocol. In the pixel front-end, the implementation of open-loop amplification led to a compact design with low noise, and the use of multiple cascode transistors helped enhance the gain of the front end~\cite{piro1}. The demonstrator's asynchronous read-out is clock-independent, transmitted through a 37-bit bus, and controlled via a 4322-bit shift register. The read-out uses a KC705, a KINTEX-7 series FPGA~\cite{lyre2}.

 \section{Micro-X-ray set-up at CERN}
 With the bunch-crossing clock of the LHC being 25 ns~\cite{bunch}, sensors and read-out architecture have to be sufficiently fast in order to match the hits with the corresponding bunch-crossing. Furthermore, nanosecond-level response time resolution is required to contribute to event reconstruction targeting future HL-LHC upgrades. The fast timing of the chip requires examination prior to its usage in a detector. Traditionally, this investigation involves reflecting a laser beam off the detector and analyzing the discrepancy between the input and output signals~\cite{hohov1, timepix4laser}. However, this laser technique proves inadequate for the MALTA2 due to significant reflection issues with the detector's front-end electronics~\cite{penneger2}, which interfere with accurate timing measurements.

\begin{figure}[ht]
    \centering
    \begin{subfigure}[b]{0.8\textwidth}
        \centering
        \includegraphics[width=\textwidth]{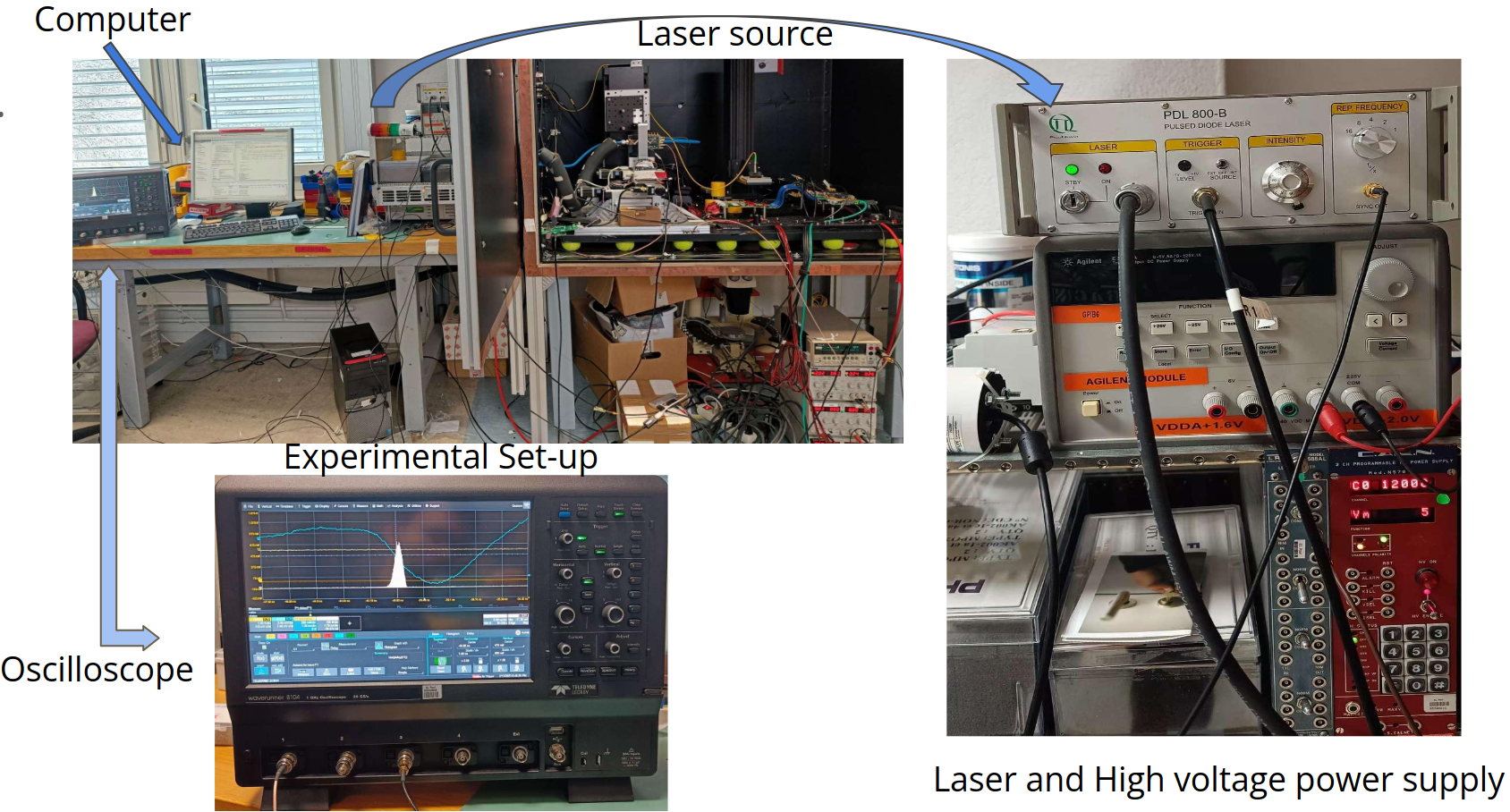}
        \caption{}
        \label{fig:setup1}
    \end{subfigure}
    
    \vspace{0.5cm}  

    \begin{subfigure}[b]{0.8\textwidth}
        \centering
        \includegraphics[width=\textwidth]{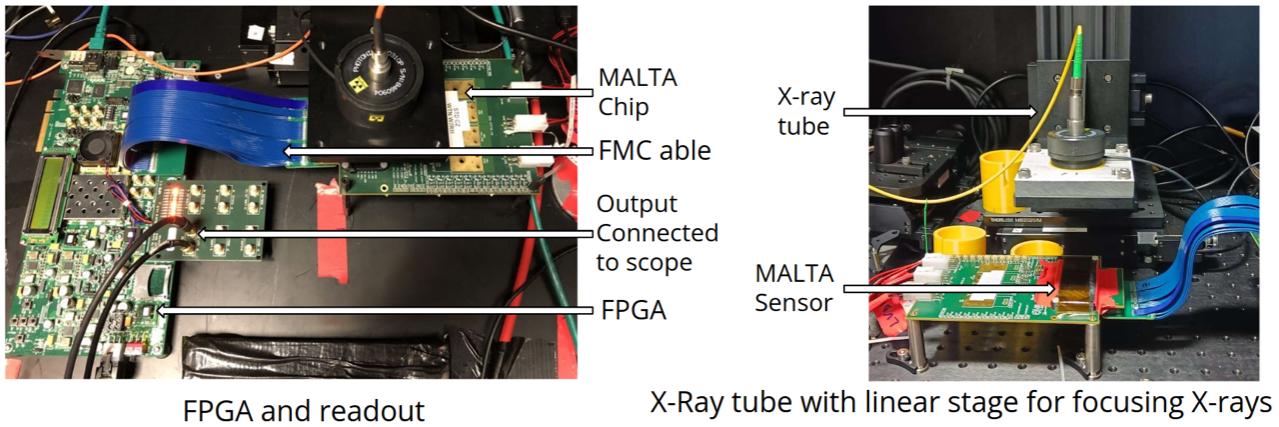}
        \caption{}
        \label{fig:setup2}
    \end{subfigure}
    \caption{\centering Figure illustrating the devices used for the study and the connections between them. In (a), the upper-left plot shows the complete set-up with a representation of a laser, a high-voltage source for X-ray generation, and an oscilloscope. In the left plot of (b), the Kintex-7 KC705 FPGA is shown, which is used to trigger the laser and also assists in signal processing. The right figure in (b) shows the MALTA sensor with the designed micro X-ray tube, held by a linear stage 10~cm above the chip.}
    \label{fig:fpga}
\end{figure}

To circumvent these challenges, we have opted for X-rays due to their ability to penetrate the metal layers of the detector without reflection. This does not require a high-end X-ray machine; instead, we have developed a custom-made micro X-ray system, as shown in Figure~\ref{fig:fpga}, which is capable of emitting X-rays in response to laser exposure. This approach employs a micro X-ray source that generates short pulses from Cu-Cr targets specifically designed for this application.

For the purpose of characterizing the timing performance of MALTA2, our triggered X-ray set-up plays an important role. Initially, the timing response of the X-ray itself is measured using an LGAD~\cite{pelle1}. The LGAD used was a CNM LGAD with a pitch of \(1.3 \times 1.3 \, \text{mm}^2\) and a gain of about 10–20, depending on the bias voltage applied~\cite{allaire}. The readout was done with discrete electronics, including a first and second stage amplifiers~\cite{allaire}. This baseline measurement serves as a reference for further comparison with MALTA2's timing response, enabling us to achieve a comprehensive understanding of the detector's timing capabilities.

\subsection{Micro-X-ray Source}

\begin{figure}[ht]
\begin{center}
\includegraphics[width=7cm]{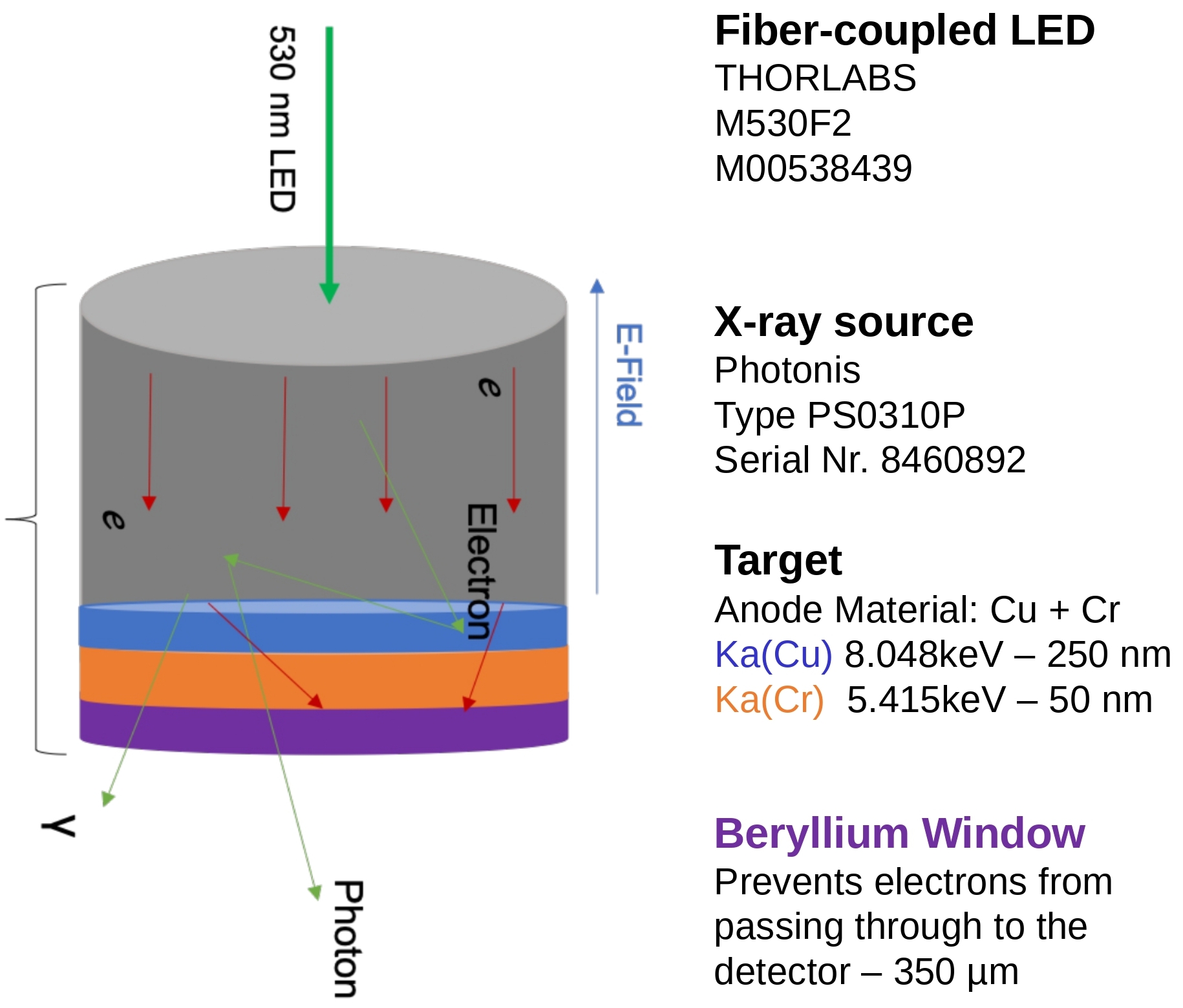}
\caption{\centering{Illustrative description of Micro X-ray source. It is connected to a 530~nm laser that falls on the cathode. The anode material is made with 250~nm copper and 50~nm chromium, which generates X-rays after interacting with accelerated electrons. Finally, there is a beryllium window that prevents electrons from passing through it.}}
\label{fig:tube}
\end{center}
\end{figure}

X-rays are generated by utilizing a 16~mm elongated tube featuring a photo-cathode and an anode composed of chromium and copper, fabricated above a Beryllium window. Electrons are emitted from the photo-cathode via the photoelectric effect initiated by a laser beam. Subsequently, these emitted electrons undergo acceleration through the application of a 12~kV electric field within the tube. The interaction of these accelerated electrons with the target material in the anode produces X-rays, including bremsstrahlung and fluorescence. The X-ray production process is illustrated in Figure~\ref{fig:tube}. The emitted X-rays exhibit a random trajectory in space. With an increased applied voltage to the X-ray tube, the number of generated photons increases due to the formation of new transition lines.

\subsection{Geant4 Simulation for X-ray tube}

\begin{figure}[h]
\begin{center}
\includegraphics[width=7.2cm]{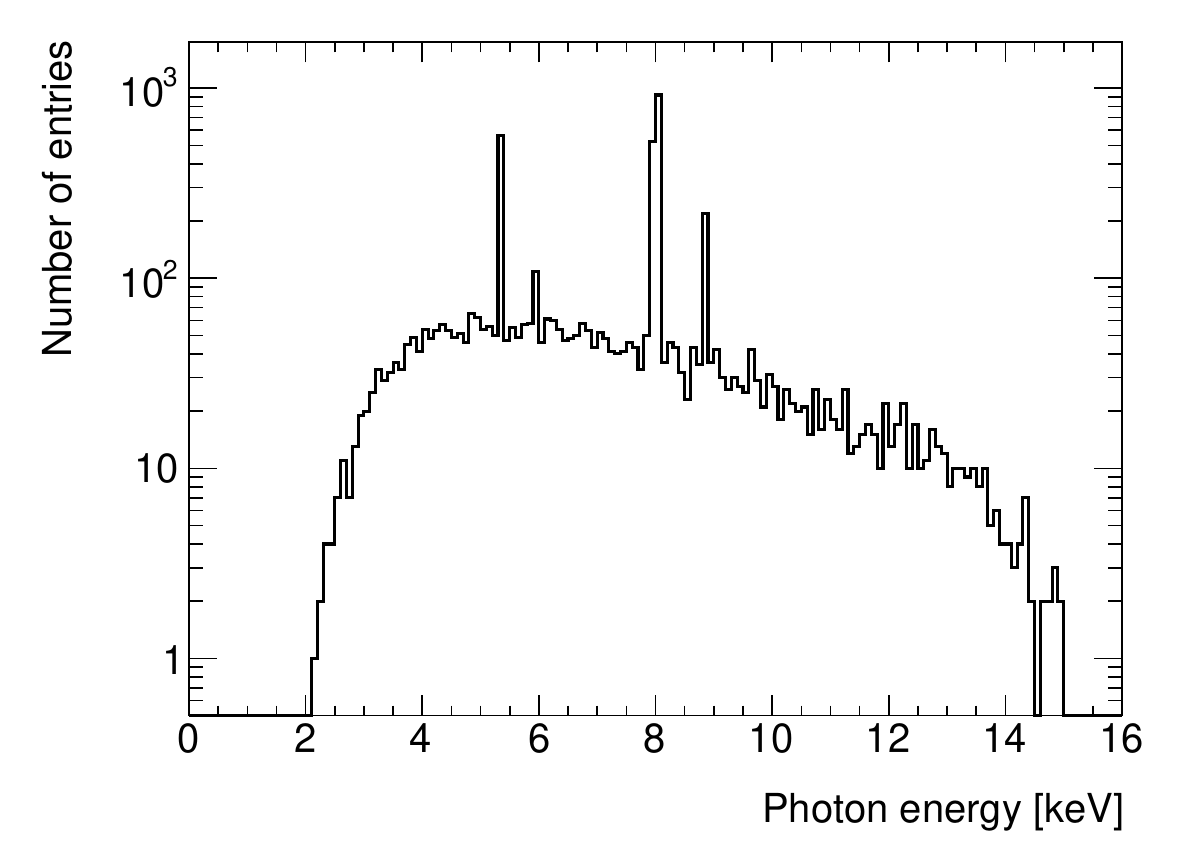}
\caption{\centering Geant4 simulation of photon energies generated from the micro X-ray tube. The x-axis represents the photon energy in keV. Four characteristic peaks are visible: two prominent peaks at 5.41 and 8.04~keV corresponding to the $K_\alpha$ lines of Cr and Cu, and two smaller peaks at 5.95 and 8.90~keV corresponding to their $K_\beta$ lines, respectively.}
\label{fig:Geant4}
\end{center}
\end{figure}

A Geant4 simulation was conducted to analyze the energy variation of photons emitted from an X-ray tube. The simulation results in Figure~\ref{fig:Geant4} show distinct peaks at 5.41 and 8.04~keV, corresponding to the characteristic $K_\alpha$ lines of chromium and copper, along with two distinct $K_\beta$ lines. Subsequently, experimental validation of these findings was undertaken using an LGAD (Low Gain Avalanche Detector) because of its high granularity and precise timing capabilities.

\section{Test Set-up with LGAD}

The LGAD, or Low-Gain Avalanche Detector~\cite{pelle1,allaire,carulla}, is a type of silicon detector used in particle physics experiments. The utilization of LGAD in this experimental phase is particularly advantageous, as it enables optimal timing resolutions~\cite{diehl1}, a crucial factor in accurately characterizing X-ray emission.

\begin{figure}[htbp]
\centering

\begin{subfigure}[b]{0.45\textwidth}
\centering
\begin{tikzpicture}[
    node distance=2.7cm, 
    node1/.style={fill=cyan!30, draw, minimum size=1.5cm},
    node2/.style={text=black},
    arrow/.style={-Stealth, thick}
]

\node[node1] (laser) {laser};
\node[node1] (oscilloscope) [below=of laser] {OSCILLO SCOPE};
\node[node1] (lgad) [right=of oscilloscope] {LGAD};
\draw [arrow] (laser) -- (lgad);
\draw [arrow] (lgad) -- (oscilloscope);
\draw [arrow] (laser) -- (oscilloscope);

\node[node2] at ($(laser)!0.5!(lgad)$) [above, yshift=0.2cm, rotate=-40] {2. Laser emission};
\node[node2] at ($(lgad)!0.5!(oscilloscope)$) [below, xshift=0.4cm, yshift=-0.2cm] {3. LGAD output};
\node[node2] at ($(oscilloscope)!0.5!(laser)$) [left, xshift=2cm, yshift=3.6cm] {1. Trigger the laser};

\end{tikzpicture}
\caption{Without using a micro X-ray source}
\label{fig:sub1}
\end{subfigure}
\hfill
\begin{subfigure}[b]{0.45\textwidth}
\centering
\begin{tikzpicture}[
    node distance=3.2cm, 
    node1/.style={fill=cyan!30, draw, minimum size=1.5cm},
    node2/.style={text=black},
    arrow/.style={-Stealth, thick}
]

\node[node1] (laser) {laser};
\node[node1] (xray) [right=of laser] {Micro X-ray};
\node[node1] (lgad) [below=of xray] {LGAD};
\node[node1] (oscilloscope) [below=of laser] {OSCILLO SCOPE};

\draw [arrow] (laser) -- (xray);
\draw [arrow] (xray) -- (lgad);
\draw [arrow] (lgad) -- (oscilloscope);
\draw [arrow] (laser) -- (oscilloscope);

\node[node2] at ($(laser)!0.5!(xray)$) [above, xshift=-0.2cm, yshift=0.2cm] {2. Laser emission};
\node[node2] at ($(xray)!0.5!(lgad)$) [right, xshift=0.4cm, yshift=-1.4cm, rotate=90] {3. X-ray Emission};
\node[node2] at ($(lgad)!0.5!(oscilloscope)$) [below, xshift=0.4cm, yshift=-0.2cm] {4. LGAD output};
\node[node2] at ($(oscilloscope)!0.5!(laser)$) [left, xshift=2cm, yshift=3.6cm] {1. Trigger the laser};

\end{tikzpicture}
\caption{With using a micro X-ray source}
\label{fig:sub2}
\end{subfigure}

\caption{\centering LGAD set-up illustrating the working flow of the experiment. In (a), the laser is focused directly on the LGAD, and in (b), an additional micro X-ray source is used. The laser emission and detection time information are collected and compared in both cases using an oscilloscope to extract individual contributions from the laser and the X-ray source.}
\label{fig:lgad}
\end{figure}
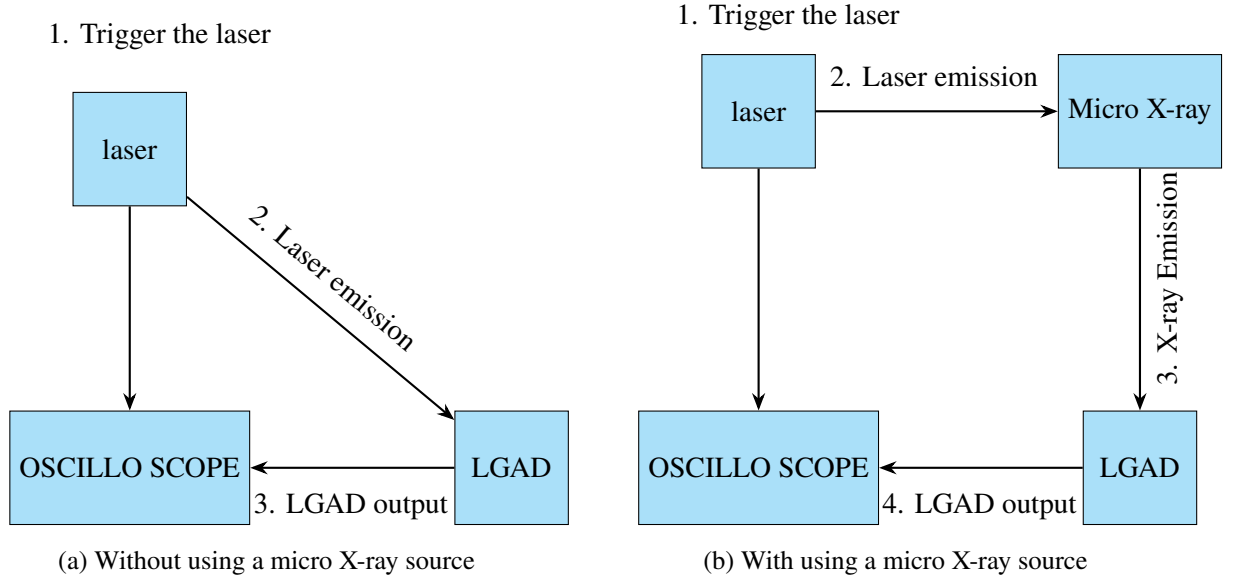

The LGAD experimental arrangement is illustrated in the accompanying schematic diagram, shown in Figure~\ref{fig:lgad}. In this set-up, laser beams are generated and guided onto a micro X-ray source, initiating the generation of X-rays. The resultant X-rays are directed toward the LGAD, and the corresponding signals are captured and recorded using an oscilloscope. Simultaneously, the LGAD signal is directly used to access the energy of the emitted photons based on the amplitude of the LGAD signal. This approach provides a direct and efficient means of evaluating the photon energy within our experimental framework.

\section{Results from LGAD}

We conducted two key studies utilizing the Low-Gain Avalanche Diode. The first study focused on characterizing the energy peaks generated by the X-ray source, which is necessary to understand the X-ray tube's performance and ensure its suitability for our applications. The second study analyzed the timing precision of the X-ray source by examining the standard deviation of its delay. For this purpose, we compared the performance of the micro X-ray source, activated by laser exposure, against a standalone laser source. These preliminary analyses are important for evaluating the contribution of the X-ray set-up to the total delay, ensuring precise timing measurements in our experimental framework.

\subsection{Characterization of Micro X-ray Tube}

\begin{figure}[ht]
\begin{center}
\includegraphics[width=10cm]{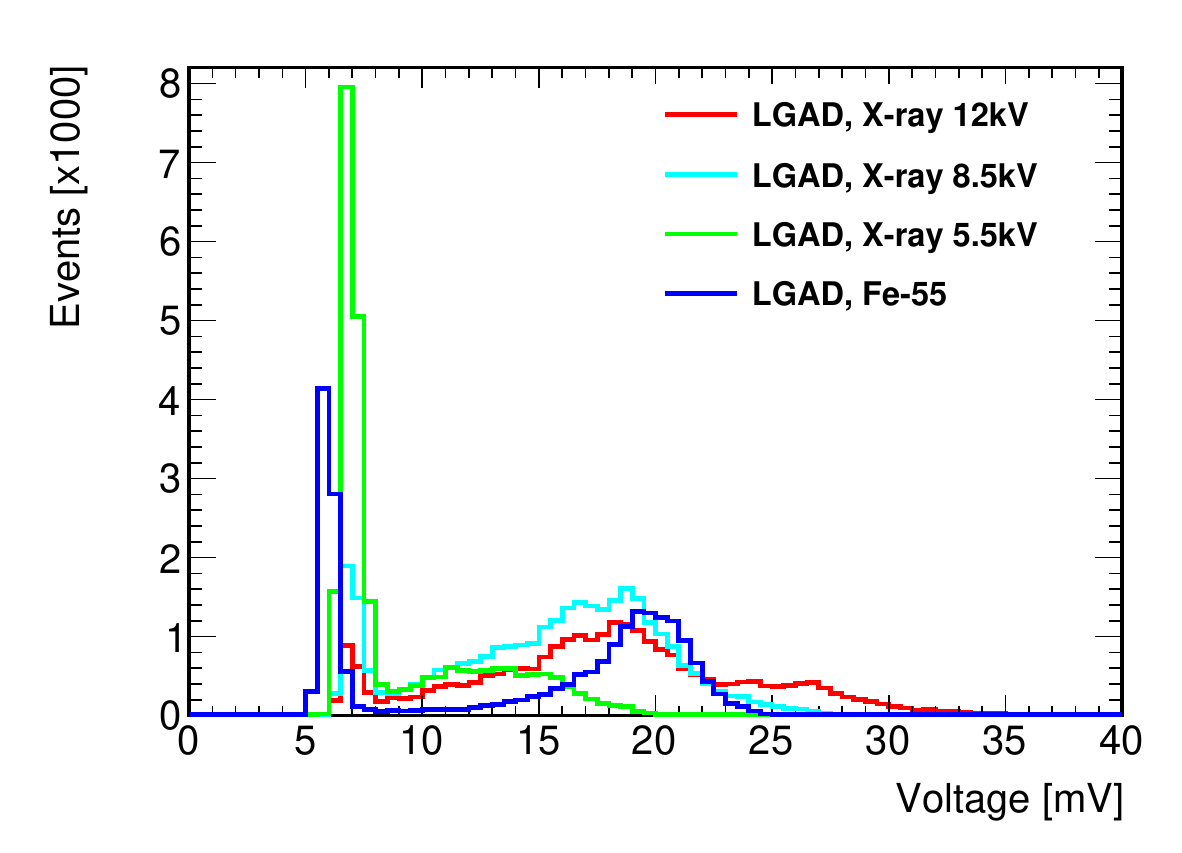}
\caption{\centering Comparison of signal amplitudes measured using the LGAD with an $^{55}\mathrm{Fe}$ source and the X-ray tube as a function of voltage applied to the LGAD. Two peaks are visible for the signal using X-rays generated at a 12~kV supply voltage, corresponding to the Cr and Cu $K_\alpha$ lines. The peaks of the red line, representing the amplitude of the signal using the X-ray tube at 12~kV, are compared with the peak from the $^{55}\mathrm{Fe}$ source (blue).}
\label{fig:peaks}
\end{center}
\end{figure}

\begin{figure}[ht]
\begin{center}
\includegraphics[width=15cm]{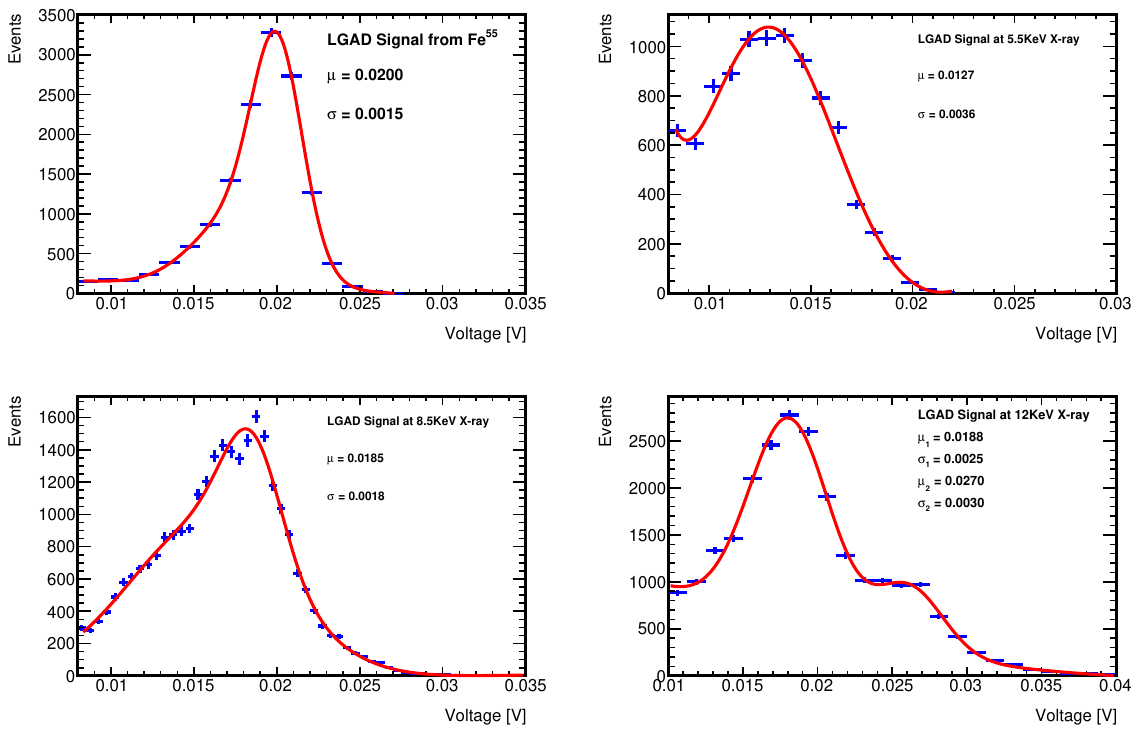}
\caption{\centering Comparison of the fitting parameters of generated signal amplitudes using the LGAD with $^{55}\mathrm{Fe}$ and X-ray sources at 5.5~kV, 8.5~kV, and 12~kV. The x-axis represents the energy of generated electrons from the LGAD using the sources, while the y-axis represents the number of events. Two peaks are visible for the signal amplitude using X-rays generated at 12~kV supply voltage. These plots are fitted using Gaussian(s) for peaks and a polynomial function to fit the initial points.}
\label{fig:fit1}
\end{center}
\end{figure}

In this study, we quantify the energy of photons using the LGAD signal, represented on the x-axis in terms of voltage and exhibiting a linear relationship with energy. For calibration, we employed a $^{55}\mathrm{Fe}$ source, which emits $K_\alpha$ X-rays with a well-known energy of 5.9~keV, thereby establishing the calibration framework. The measured energy from the alpha decay was subsequently compared with the energies of X-rays generated at voltages of 5.5~kV,\, 8.5~kV,\, and 12~kV.\, Notably, the X-rays produced under a 12~kV voltage setting revealed distinct peaks at 5.4~keV and 8~keV, which correspond closely to the theoretical $K_\alpha$ emission lines of chromium and copper, respectively. These peaks occur because the incident electron energy at 12~kV exceeds the K-shell binding energies of these elements, allowing characteristic X-ray lines to appear alongside the bremsstrahlung continuum. As the energy of the X-ray photon increases, the emitted electron gains more energy, resulting in a higher signal voltage. A comparison of peaks is shown in Figure~\ref{fig:peaks}. The peak is parametrized using Gaussian and polynomial functions, as shown in Figure~\ref{fig:fit1}. The results from the fit are summarized in Table~\ref{tab:teb2}, considering the proportional values of output signal voltages with X-ray energy, which match the theoretical values of $K_\alpha$ lines from chromium and copper.

\begin{table}[h]
\begin{center}
\caption{Summary of the fit results}
\label{tab:teb2}
\begin{tabular}{p{6cm}ccc}
\hline\hline
\textbf{Peak}             & \makecell{\textbf{Value} \\ \textbf{(in V)} }   & \makecell{\textbf{Energy} \\ \textbf{(Theoretical value)}}    & \makecell{\textbf{Energy} \\ \textbf{(Observed value)}}\\
\hline
Fe source   & 0.02             & 5.9 keV    & 5.9 keV \\
Peak from 8.5 keV Cr     & 0.0185          & 5.41 keV      & 5.45 keV \\
Peak from 12 keV Cr      & 0.0188          & 5.41 keV     & 5.54 keV \\
Peak from 12 keV Cu      & 0.0270          & 8.04 keV     & 7.96 keV \\
\hline\hline
\end{tabular}
\end{center}
\end{table}

\subsection{Timing performance of LGAD}

\begin{figure}[ht]
    \centering
    
    \begin{subfigure}[b]{0.3\textwidth}
        \includegraphics[width=\textwidth]{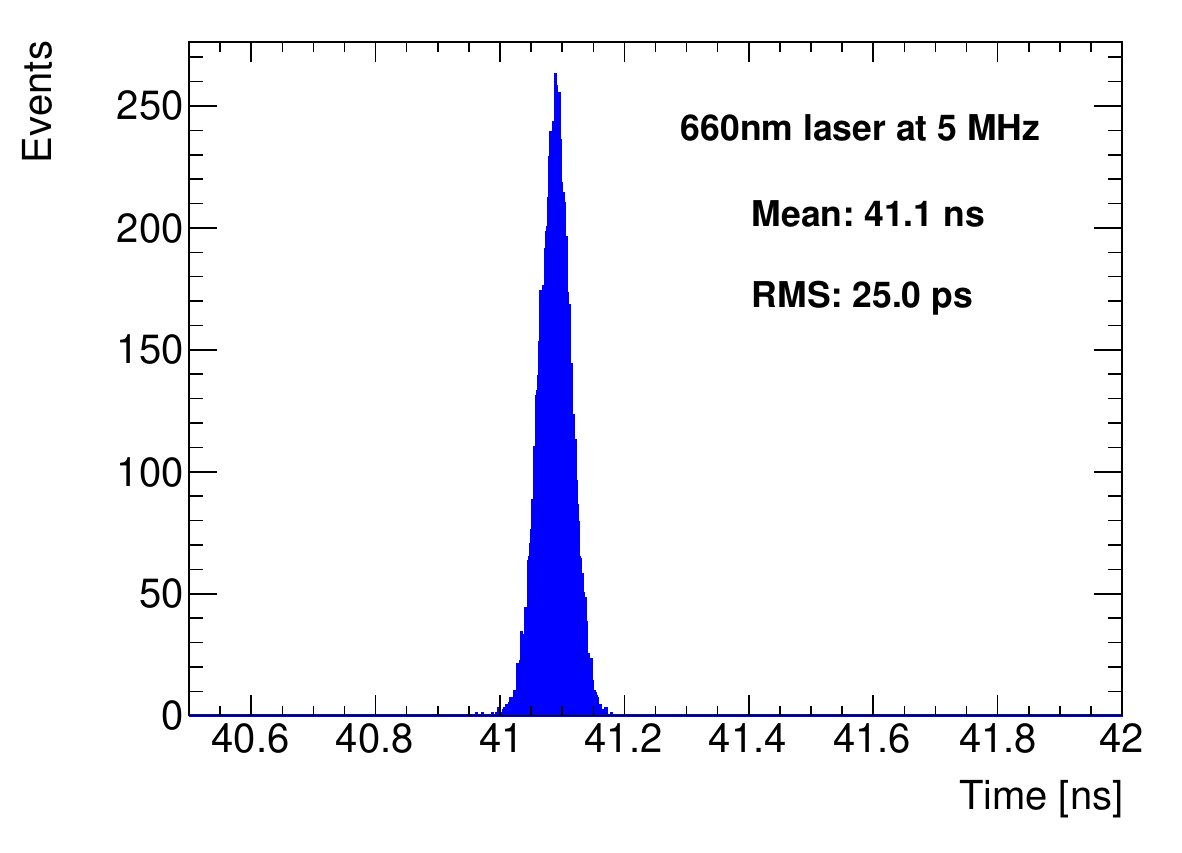}
        \caption{\centering{}}
        \label{fig:subfigs1}
    \end{subfigure}
    \hspace{0.02\textwidth}
    \begin{subfigure}[b]{0.3\textwidth}
        \includegraphics[width=\textwidth]{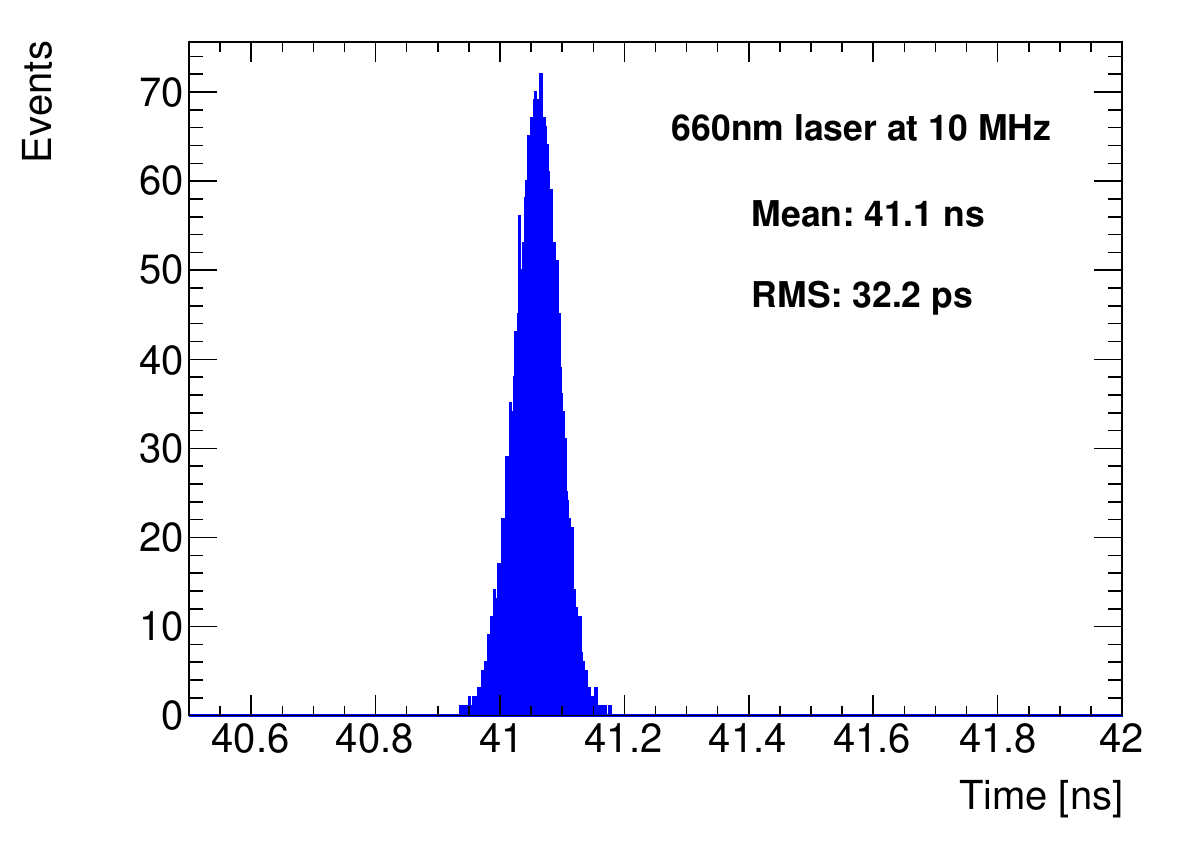}
        \caption{\centering{}}
        \label{fig:subfigs2}
    \end{subfigure}
    \hspace{0.02\textwidth}
    \begin{subfigure}[b]{0.3\textwidth}
        \includegraphics[width=\textwidth]{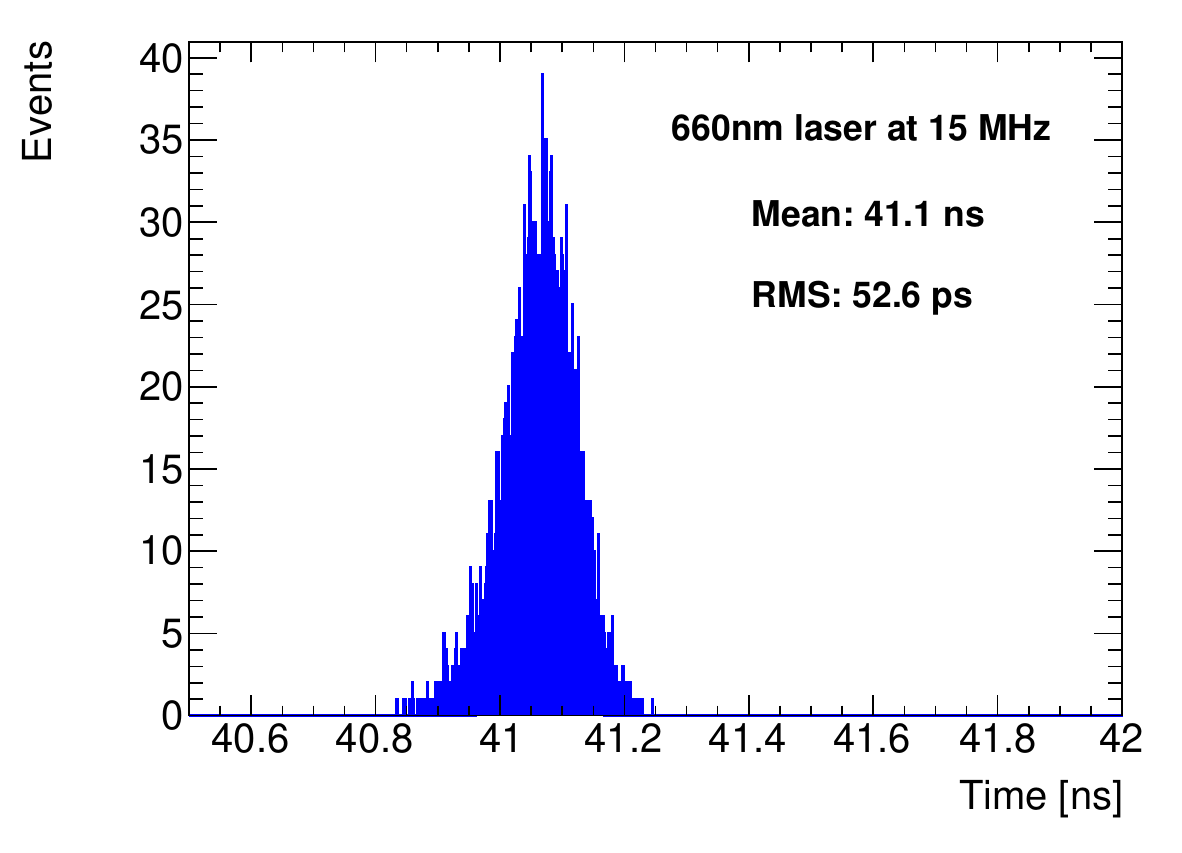}
        \caption{\centering{}}
        \label{fig:subfigs3}
    \end{subfigure}
    
    \vspace{0.5cm}
    
    \begin{subfigure}[b]{0.3\textwidth}
        \includegraphics[width=\textwidth]{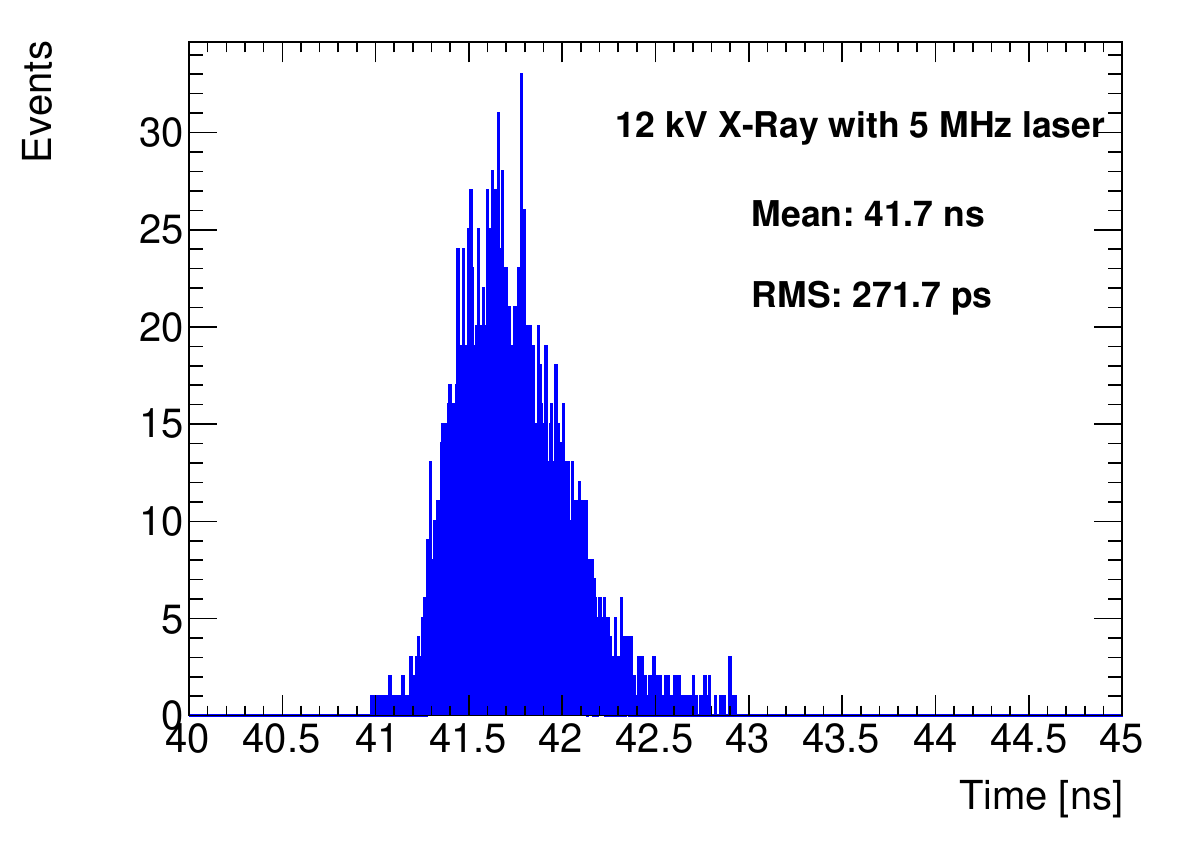}
        \caption{\centering{}}
        \label{fig:subfigs4}
    \end{subfigure}
    \hspace{0.02\textwidth}
    \begin{subfigure}[b]{0.3\textwidth}
        \includegraphics[width=\textwidth]{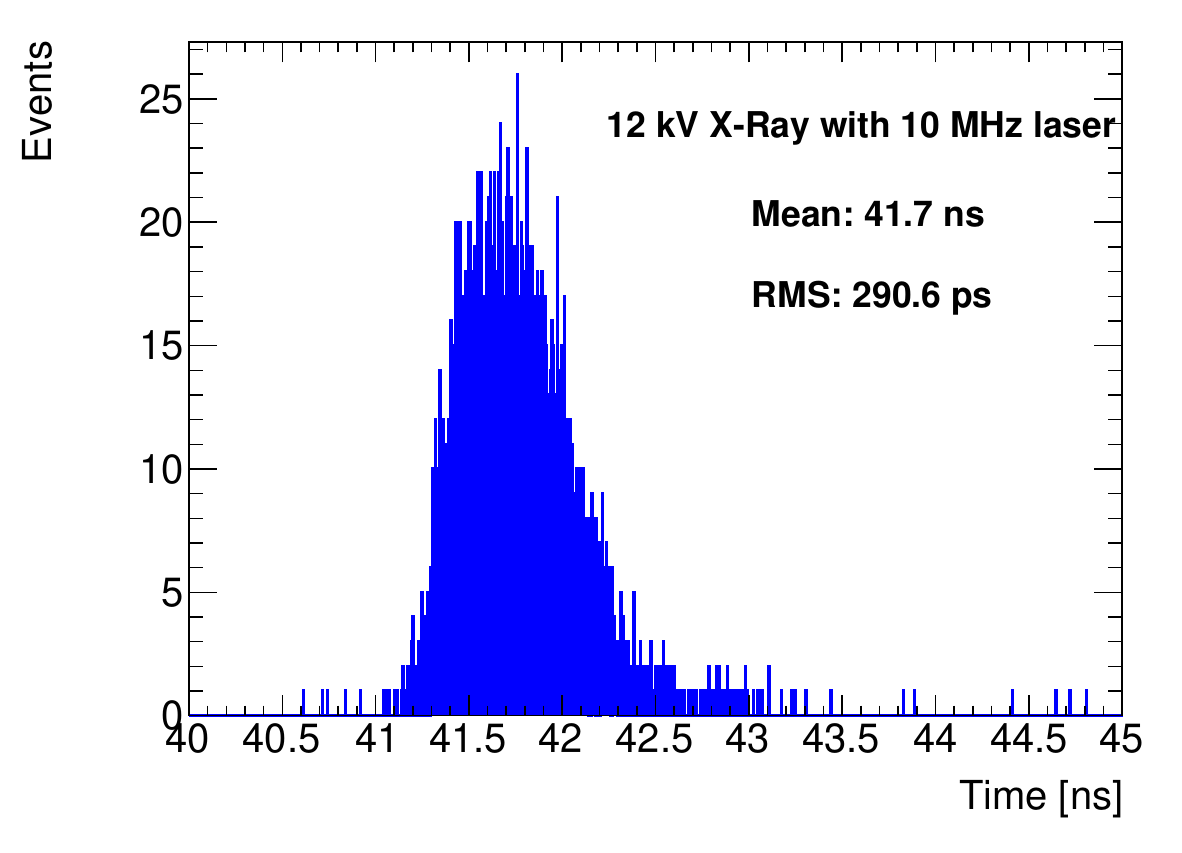}
        \caption{\centering{}}
        \label{fig:subfigs5} 
    \end{subfigure}
    \hspace{0.02\textwidth} 
    \begin{subfigure}[b]{0.3\textwidth}
        \includegraphics[width=\textwidth]{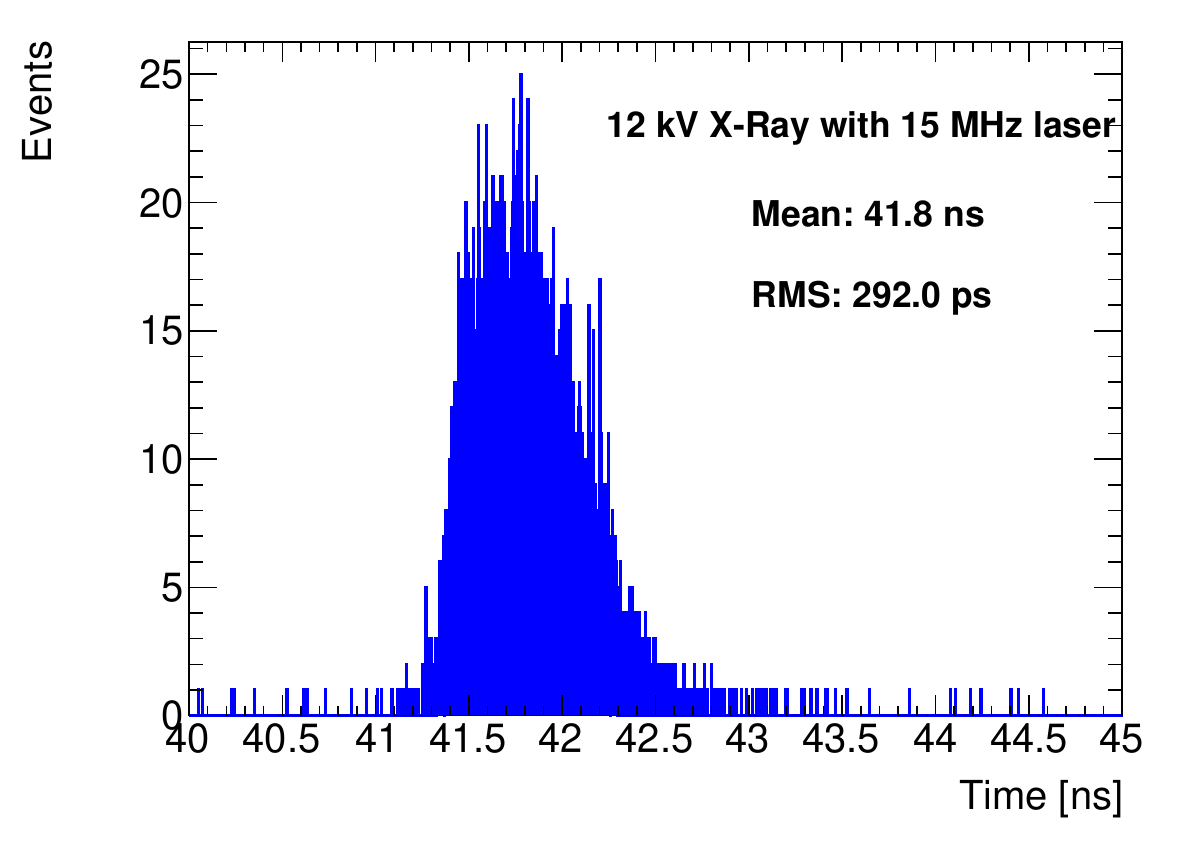}
        \caption{\centering{}}
        \label{fig:subfigs6}
    \end{subfigure}
     
\caption{\centering Comparison of the timing delay between the laser trigger and LGAD signal detection using the laser and X-rays. The laser frequency is varied from 5 to 15~MHz, and the RMS of the delay is compared without (Figure~\ref{fig:sub1}) and with (Figure~\ref{fig:sub2}) the micro X-ray source. The RMS of the timing delay using the micro X-ray tube with the laser is found to be approximately 250~ps higher than the RMS of the timing delay without the micro X-ray tube, indicating the contribution of the micro X-ray tube to the RMS of the delay.}
\label{fig:laser}
\end{figure}

In this investigation, the LGAD was characterized using a laser beam, yielding timing resolutions in the range of 25–50~ps, where the resolution includes contributions from both the LGAD and the laser source. Subsequently, the entire system was employed to generate X-rays through laser irradiation. From the analysis of the X-ray signals, the timing resolution was determined to be approximately 270–300~ps across varying laser frequencies, as shown in Figure~\ref{fig:laser}. This evaluation allowed us to infer that the intrinsic timing resolution of the micro X-ray source is estimated to be in the range of 269–295~ps (\(\sqrt{270^2 - 25^2}\)–\(\sqrt{300^2 - 50^2}\)~ps), demonstrating its efficiency and reliability in temporal measurements. These findings highlight the potential of our system for applications requiring precise timing resolution in X-ray generation and detection on the order of 100~ps.

\section{Measurements with MALTA and MALTA2}

\subsection{Measurements Using MALTA Sample}

\begin{figure}[htbp]
\centering
\begin{tikzpicture}[
    node distance=3.8cm, 
    node1/.style={fill=cyan!30, draw, minimum size=1.5cm},
    node2/.style={text=black},
    arrow/.style={-Stealth, thick}
]

\node[node1] (sw) {SW};
\node[node1] (fpga) [right=of sw] {FPGA};
\node[node1] (laser) [right=of fpga] {Laser};
\node[node1] (xray) [below=of laser] {Micro X-ray};
\node[node1] (sensor) [below=of fpga] {MALTA Sensor};
\node[node1] (oscilloscope) [below left=of fpga] {Oscilloscope};

\draw [arrow] (sw) -- (fpga);
\draw [arrow] (fpga) -- (laser);
\draw [arrow] (laser) -- (xray);
\draw [arrow] (xray) -- (sensor);
\draw [arrow] (sensor) -- (fpga);
\draw [arrow] (fpga) -- (oscilloscope);

\node[node2] at ($(sw)!0.5!(fpga)$) [above, yshift=0.2cm] {1. Trigger the FPGA};
\node[node2] at ($(fpga)!0.5!(laser)$) [above, yshift=0.2cm] {2. Trigger the laser};
\node[node2] at ($(laser)!0.5!(xray)$) [right, xshift=0.4cm, yshift=-1.4cm, rotate=90] {3. Laser emission};
\node[node2] at ($(xray)!0.5!(sensor)$) [above, xshift=0.1cm, yshift=0.2cm] {4. X-ray emission};
\node[node2] at ($(sensor)!0.5!(fpga)$) [right, xshift=0.4cm, yshift=-1.4cm, rotate=90] {5. Hits to FPGA};
\node[node2] at ($(fpga)!0.5!(oscilloscope)$) [left, xshift=0.3cm]{\parbox{2cm}{6. output\\2. input}};
\end{tikzpicture}
\caption{\centering{Schematic diagram of the MALTA testing layout illustrating the working flow of the experimental set-up. Upon instruction, the FPGA triggers the laser and sends the actual triggering time to the oscilloscope. The laser falls on the X-ray tube, resulting in X-ray emission. The sensor detects the X-ray exposure and sends the detection time to the oscilloscope.}}
\label{fig:malta}
\end{figure}
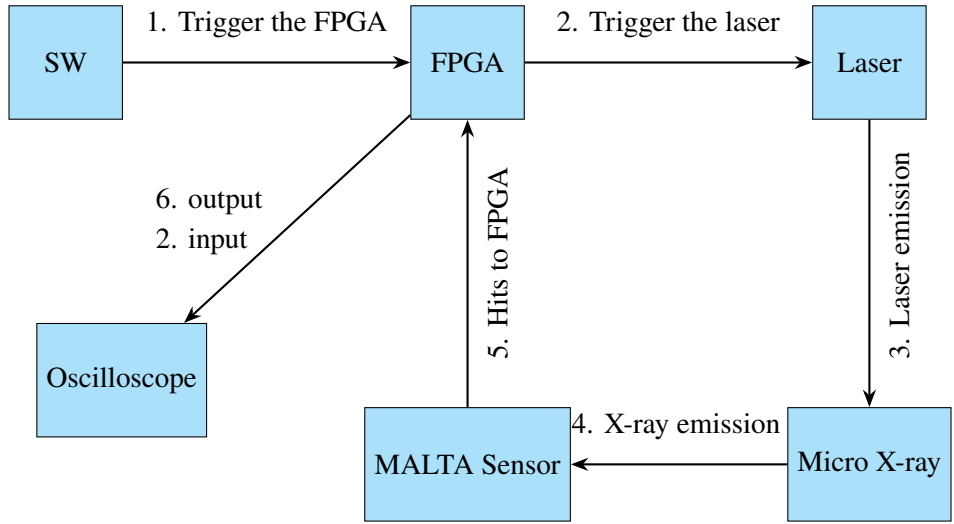

In contrast to the LGAD set-up, the MALTA system employs an internal trigger from the FPGA to initiate the laser, which is subsequently directed toward the X-ray tube for X-ray emission, as described in Figure~\ref{fig:malta}. The operational sequence involves triggering the FPGA using software, whereby the KINTEX-7 FPGA activates the laser source. As the laser beam interacts with the X-ray tube, X-rays are emitted and directed toward the MALTA sensor (W11R12, a 100~nm thick sensor with a p-type Czochralski substrate and an extra deep p-well implant). The timing information is gathered within the FPGA and transmitted to the oscilloscope, which then compares the timing data to determine the delay. This delay information is necessary for analyzing the temporal characteristics of MALTA, where the effect of the previously measured micro X-ray source (approximately 300~ps) can be adjusted. Notably, the timing performance of MALTA is observed to operate on a nanosecond scale, showcasing its capability for precise measurements.

\subsection{Measurements Using MALTA2 Sample}

\begin{figure}[ht]
\begin{center}
\includegraphics[width=10cm]{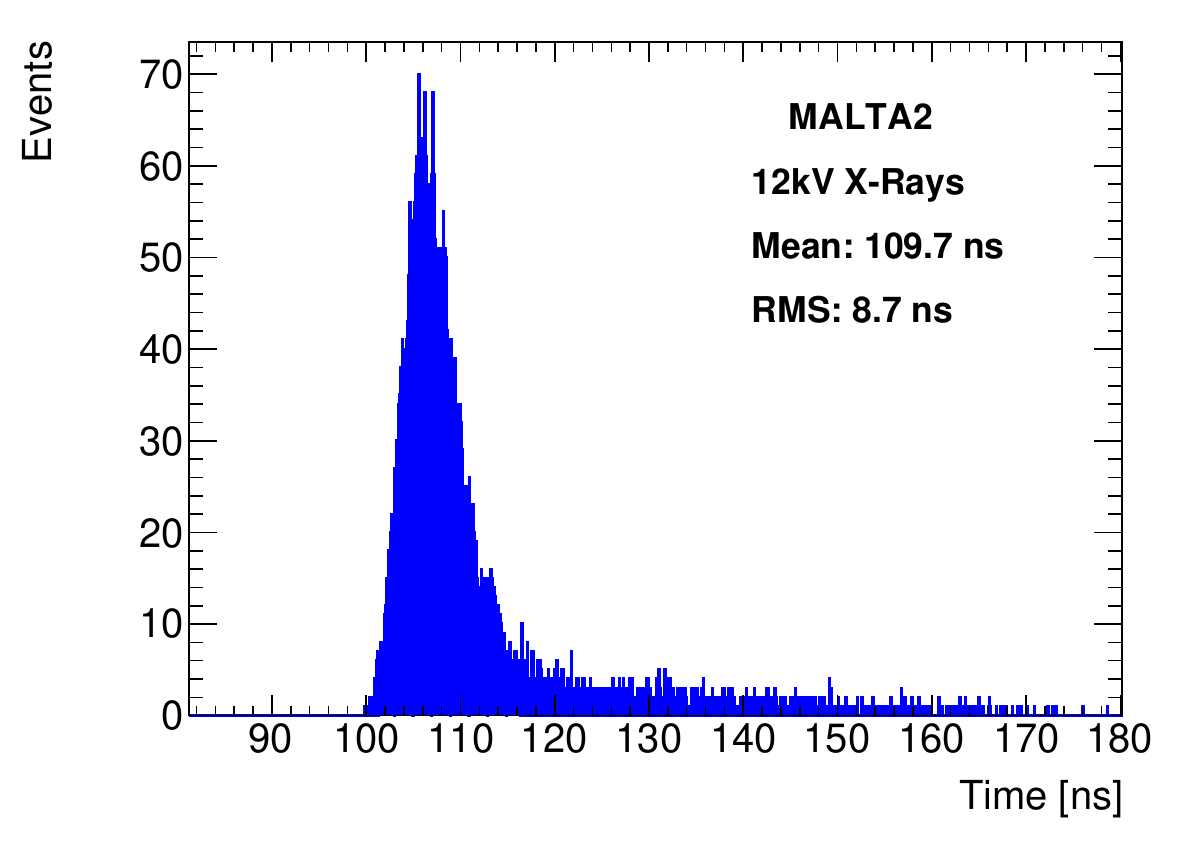}
\caption{\centering Delay between the input and output signals using 12~kV X-rays on the MALTA2 sample. Contributions to the tail part come from corner pixel hits, front-end electronics, as well as comparison errors from the oscilloscope.}
\label{fig:malta2}
\end{center}
\end{figure}

While the set-up for the MALTA2 chip mirrors that of MALTA, distinctions in the MALTA2 configuration necessitate corresponding alterations in connection and firmware. Specifically, the default FPGA firmware for MALTA2 is modified to receive signals from the computer, subsequently triggering the laser and capturing the timing information when X-rays interact with the sensor. These firmware modifications are crucial to facilitate the transmission of signals to the oscilloscope, ensuring proper data collection and analysis. The timing study of MALTA2, an epitaxial chip with an extra deep p-well implant, is conducted over a range of high voltages with different threshold values. The delay for 12~kV operation can be seen in Figure~\ref{fig:malta2}.

 \section{Results}

\begin{figure}[ht]
    \centering
    \begin{subfigure}[b]{0.8\textwidth}
        \centering
        \includegraphics[width=\textwidth]{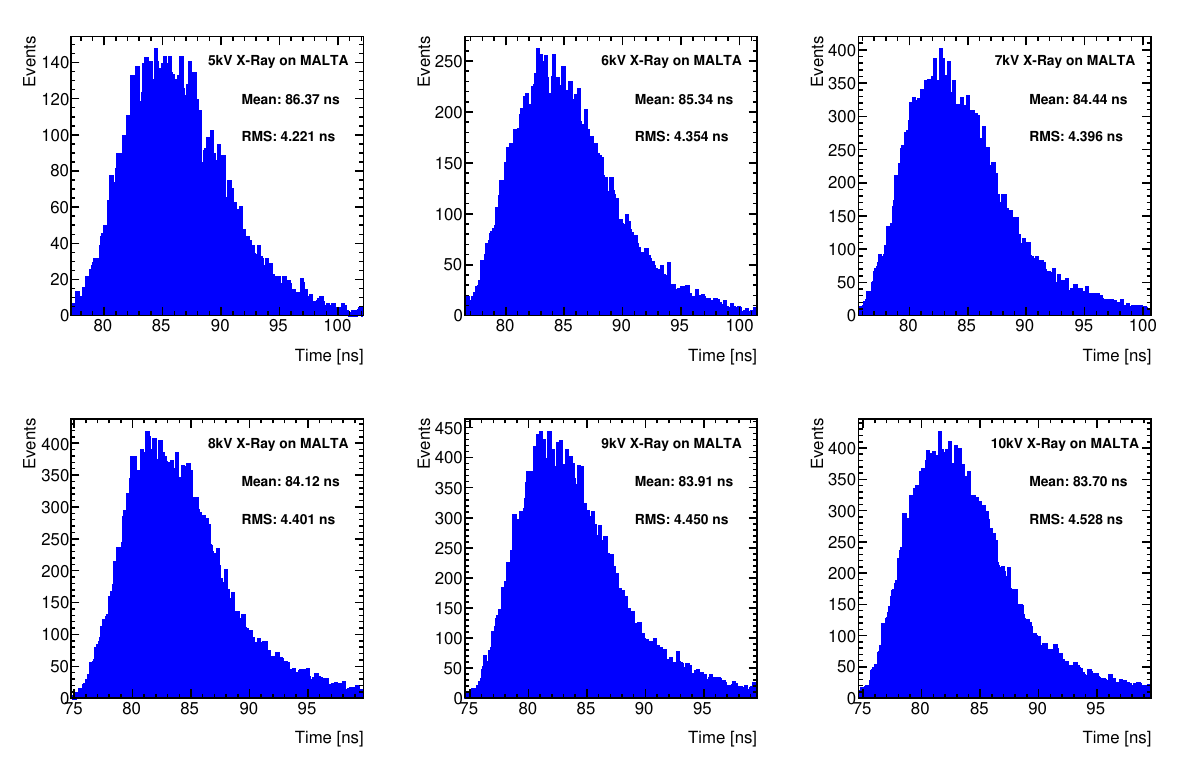}
        \caption{}
        \label{fig:MALTA_range}
    \end{subfigure}
    
    \vspace{0.5cm}  

    \begin{subfigure}[b]{0.8\textwidth}
        \centering
        \includegraphics[width=\textwidth]{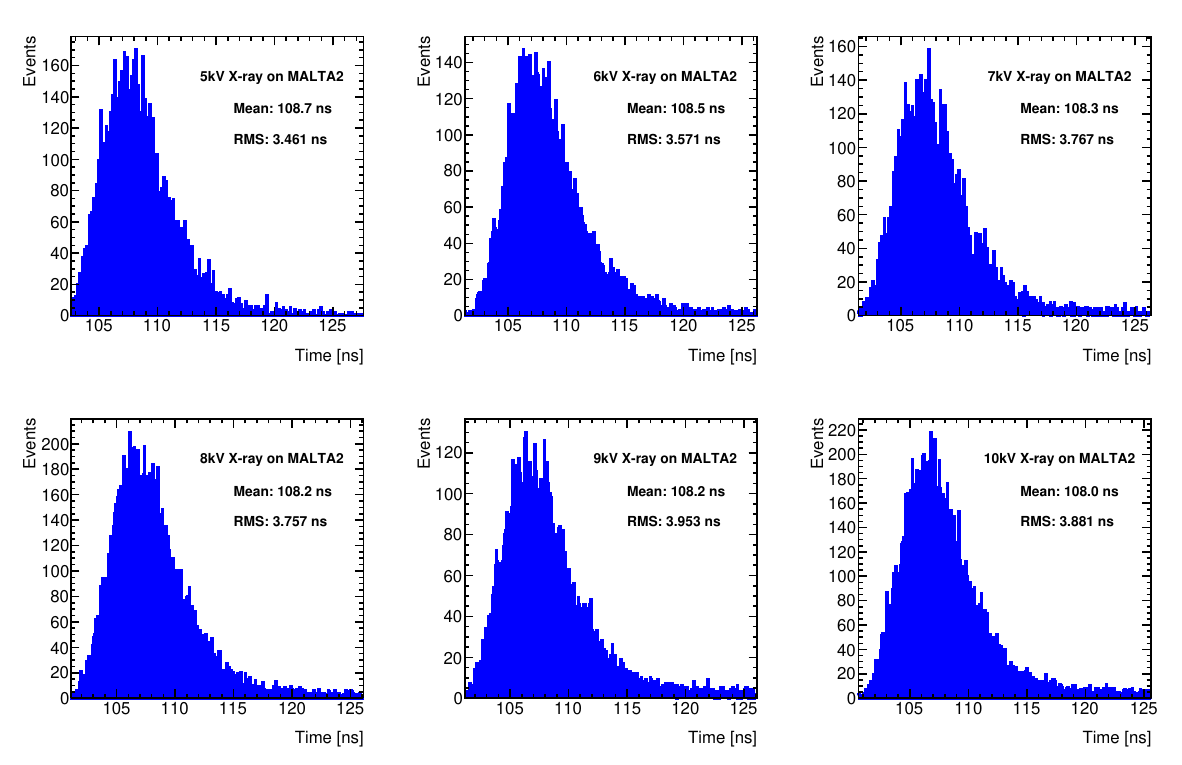}
        \caption{}
        \label{fig:MALTA2_range}
    \end{subfigure}
    
   \caption{\centering Comparison of the timing delay RMS for MALTA and MALTA2 samples, evaluated within the first 25~ns. Subfigures in (a) show the timing distribution for MALTA, and Subfigures in (b) show the timing distribution for MALTA2. The x-axis represents the input voltage applied to the micro X-ray tube, while the y-axis indicates the number of hits within the first 25~ns. The mean and RMS values of the respective distributions are displayed in each panel. Timing delay contributions from the measurement set-up are included for both MALTA and MALTA2 samples.}
\label{fig:peakcomp}
\end{figure}

\begin{figure}[ht]
\begin{center}
\includegraphics[width=10cm]{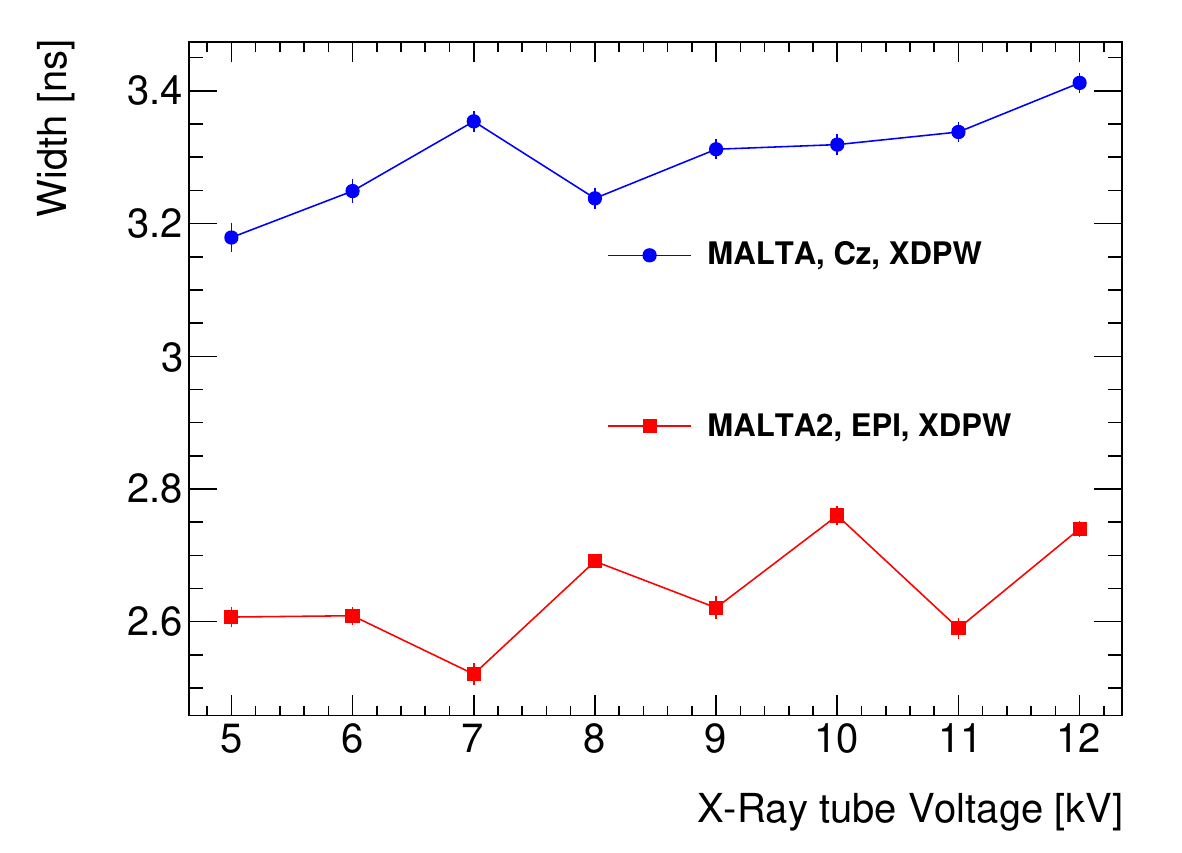}
\caption{\centering Comparison between the standard deviation of the timing delay of MALTA and MALTA2 samples using the Gaussian fit method. The peaks are fitted with a single Gaussian, and the width of the Gaussian fit is compared for MALTA and MALTA2 at different X-ray tube voltages. The x-axis shows the input voltage applied to the micro X-ray tube, and the y-axis represents the standard deviation of the Gaussian fit. Timing delay contributions from the experimental set-up are included for both the MALTA and MALTA2 samples.}
\label{fig:Gaussianfit}
\end{center}
\end{figure}

A distinct and extended tail is evident in both the MALTA and MALTA2 datasets. It is crucial to emphasize that this phenomenon is not solely attributed to timing differences. Rather, the presence of such pronounced tails can largely be attributed to errors introduced by the oscilloscope. Additional contributions to the tail arise from the MALTA front-end electronics. The underlying mechanism involves the acquisition of two distinct signals from the FPGA in the oscilloscope—one originating from the trigger and the other from the MALTA2 output. These signals are characterized by multiple peaks rather than a single peak. Our approach to determining the timing difference involves selecting the corresponding peaks; however, occasional discrepancies result in the acceptance of the timing associated with the subsequent peak rather than the intended one, leading to the augmentation of data points in the far-tail region.

To mitigate this effect, we adopted two distinct methodologies. Initially, we calculated the root mean square (RMS) value by excluding the prolonged tail and focusing solely on the initial 25~ns (bunch-crossing time for the LHC), as shown in Figure~\ref{fig:peakcomp}. From this plot, it is evident that the RMS increases with the applied voltage to the X-ray tube, primarily due to the involvement of a higher number of transition lines in Cu and Cr at higher applied voltage.

Subsequently, an alternative approach involved a comparative analysis of the peak width for both MALTA2 and MALTA datasets. Although these peaks do not perfectly match Gaussian fits, our analysis considered only single Gaussian fits for general applicability to derive approximate values. We then compared the width values of MALTA and MALTA2 from the fit at different applied voltages to the X-ray tube.

From Figure~\ref{fig:Gaussianfit}, we find that the RMS of the peak is approximately 2.6~ns for MALTA2, with a best value of 2.5~ns. This value includes contributions from the setup (\(\sim 300\)~ps), jitter from the FPGA (\(\sim 160\)~ps), and jitter from the laser source (up to 50~ps), all of which affect the final measurement. After correcting for these contributions, the RMS of the fast timing is calculated as
\[
\sqrt{2600^2 - 300^2 - 160^2 - 50^2} \ \text{ps} \approx 2570~\text{ps} \ (2.57~\text{ns}),
\]
with the best measured value of 2.5~ns for the MALTA2 epitaxial sample.

Similarly, for MALTA, the RMS of the peak is found to be approximately 3.3~ns, with a best value of 3.2~ns. After applying the corrections, the fast timing resolution lies within 3.18–3.28~ns. In this case, we observe consistent values (unlike in Figure~\ref{fig:peakcomp}), as the contribution from photons generated at higher transition levels is reduced. Furthermore, considering that the intrinsic timing resolution of the MALTA sensor is predicted to be less than\(\sim 500\)~ps \cite{piro1}, the RMS values observed at different X-ray tube voltages are compatible within \(1\sigma\). This provides strong evidence of the consistent Gaussian width of the peak distributions. In both cases, MALTA2 demonstrates significantly better timing precision compared to MALTA.

From test-beam data collected between 2021 and 2023, the best timing resolution achieved for MALTA2 with a Czochralski substrate at low threshold settings was 1.7~ns using a cluster size of two pixels~\cite{milou2}. In contrast, our measurements were performed on the full chip without clustering, yet we obtained a comparable timing resolution in laboratory conditions without the need for high-intensity beams. This highlights the practicality of our approach and its potential usefulness for future measurements.

\section{Conclusion}

This study demonstrates the feasibility of using a laboratory-based micro X-ray tube for timing measurements, providing a convenient alternative to test beam campaigns. The study focused on investigating the delay time in MALTA and MALTA2 samples. Given the limitations of conventional laser techniques, a micro X-ray source comprising copper and chromium layers was employed to generate X-rays. The use of a Low-Gain Avalanche Detector (LGAD) enabled detailed characterization of the micro X-ray source, where the observed $K_\alpha$ energy peaks were consistent with theoretical values. Our findings show that the micro X-ray source contributes approximately 300~ps to the overall delay time, while the FPGA adds \(\sim 160\)~ps and the laser jitter contributes up to 50~ps. After accounting for these effects, the RMS delay time is found to be \(\sim 3.3\)~ns (best value of 3.2~ns) for MALTA and \(\sim 2.6\)~ns (best value of 2.5~ns) for MALTA2. This confirms the improved performance of the second generation of MALTA pixel detectors over the first, providing strong evidence of their suitability for future collider experiments.

\acknowledgments
This project has received funding from the European Unions Horizon 2020 Research and Innovation programme under Grant Agreement numbers: 101004761 (AIDAinnova), 675587 (STREAM), 654168 (AIDA-2020). Also, we thank Science and Engineering Research Board, India, and the Ministry of Human Resource Development, India, for their funding support.

\end{document}